\documentstyle[12pt]{article}
\begin{document}
\baselineskip 16pt
\begin{titlepage}

\begin{center}

{\bf Dynamics of Phase Behavior of a Polymer Blend Under Shear Flow}

\vfill

Tao Sun$^{1}$,  Anna C. Balazs$^{1}$, and David Jasnow$^{2}$ 

\vspace{1cm}

{$^{1}$Chemical and Petroleum Engineering Department,
University of Pittsburgh, \\ Pittsburgh, PA, 15261.
$^{2}$Department of Physics and Astronomy, \\
University of Pittsburgh, PA, 15260}
\vspace{1cm}

\end{center}

\vfill

We study the dynamics of the phase behavior of a polymer blend in the 
presence of shear flow.
By adopting a two fluid picture and using a
generalization of the concept of
material derivative, we construct kinetic equations that describe
the phase behavior of polymer blends in the presence of external flow.
A phenomenological form for the shear modulus for the blend is proposed. 
The study indicates that a nonlinear dependence of the shear modulus
of the blend on the volume fraction of one of the species is crucial 
for a shift in the stability line to be induced by shear flow. 

\vfill

\vfill

\end{titlepage}

\vspace{0.5cm}

{\bf I. INTRODUCTION}

\vspace{0.5cm}

The dynamics of the phase behavior in polymer mixtures under external
flow fields has aroused great interest over the last two
decades~\cite{Strate,Rangel,Han,Mazich,Helfand,Doi0,Doi1,Onuki,Milner,tao}.
The motivation for these studies is twofold.
First, the effect of viscoelasticity on the phase behavior of polymer
mixtures can be directly detected for some macroscopic flows. 
Secondly, many industrial processes, such as extrusion and painting
processes, generate shear flow fields in polymer solutions and melts.
In order gain further insight into non-equilibrium phase transitions and
to optimize these industrial processes, one must understand
the phase behavior of polymer mixtures in the presence of a flow field.
Experimentally, a number of groups have reported that the phase behavior
of polymer mixtures can be dramatically changed by macroscopic flow fields.
In particular, for polymer-solvent mixtures in the presence of a shear
flow, a greatly enhanced turbidity has been observed  
at temperatures much higher than the equilibrium critical 
temperature~\cite{Strate,Rangel,Han}.
To study the mechanism of the observed phenomenon in polymer solutions, 
a number of theoretical
efforts have been made~\cite{Helfand,Onuki,Milner,tao}, and it is now 
understood that a nonlinear concentration dependence of the shear modulus 
is crucial for an upward shift of the phase separation 
temperature~\cite{tao}. 
The temperature shift is proportional to the square of the shear 
strength in the regime of weak shear~\cite{Onuki,tao}.

Although some experiments have been carried out for polymer blends under
external flow~\cite{Mazich}, theoretically, one knows very little 
about the phase behavior of this system.
Doi and Onuki first discussed the Langevin equations 
describing the dynamics of phase separation of a polymer blend~\cite{Doi1}.
However, to our knowledge,
the full consequences of the equations have not been explored.
Moreover, the original approach employed in the 
ground-breaking work of Doi and Onuki
prevents these kinetic equations from reducing to the
polymer-solution case.
Here, we present an expanded derivation, in which the difference
between the Lagrangian and Eulerian descriptions has been taken into
account. 
Using a phenomenological form for the shear modulus of a polymer blend,
we can carry out a linear stability analysis for the model and find that the 
equilibrium spinodal line can be  shifted in a complicated fashion
by the shear flow.  
The purpose of this paper is to report on these studies and
describe the relevant techniques in detail.

The system consists of two kinds of polymers with different degrees 
of polymerization, $N_A$ and $N_B$.
The volume fraction of polymer $A$ at space-time point $(r,t)$ is 
denoted by $\phi_A (r,t) = \phi (r,t)$, and the volume fraction 
of polymer $B$ is then $\phi_B (r,t) = 1-\phi (r,t)$.
We make the assumption that monomers of both species have the same specific 
volume, which can be expressed as~\cite{Doi0,Doi1,Onuki}
\begin{eqnarray}
\frac{\rho_{A} (r,t)}{\phi(r,t)} & = &
\frac{\rho_{B} (r,t)}{1-\phi(r,t)} \ =\ \rho\ ,
\label{assumption-1}
\end{eqnarray}
where $\rho_A$ and $\rho_B$ are the respective mass densities of
polymers $A$ and $B$ and $\rho$ is the total mass density, which is 
a constant.
This assumption is consistent with the
incompressibility of the system.
In the two-fluid picture~\cite{Doi0,Doi1,Onuki,Milner}, the two 
species are moving with different velocities, so that both bulk flow 
of the fluid and mutual diffusion of the two species, 
accompanied by chain deformation, take place simultaneously.
Our task is to study the effect of flow and 
chain deformation on the dynamics of the phase behavior of the system. 

The paper is organized as follows.
In Sec.~II, we first discuss appropriate material derivatives
and then derive the kinetic equations for the fluid velocity $v$ and
volume fraction $\phi$.
In Sec.~III, we perform a linear stability analysis  of the model, 
from which the effect of flow on the domain of linear stability
can be obtained. Finally, our conclusions are briefly summarized 
in Sec.~IV.

\vspace{0.5cm}

{\bf II. THE MODEL }

A polymer blend is a viscoelastic system,
sharing features of an elastic continuum and a viscous 
fluid.
The system is elastic, but it only has a ``faded'' memory.
The system is viscous, but it can bear deformation on some time scales.
To derive the kinetic equations for such a system, one must
call on familiar methods in studies of deformable media, as well as
techniques for viscous fluids. 

\vspace{0.3cm}

{\bf A. Material Derivatives}

\vspace{0.3cm}

In continuum mechanics, one uses both Lagrangian and Eulerian 
coordinates to describe the motion of a material 
element~\cite{Reismann,Mal}.
The Lagrangian coordinates, which can be denoted by $a=\{a_i\},i=1,2,3$, 
are used to label the material elements or ``particles'' in a 
reference configuration 
(usually the undeformed state), while the Eulerian coordinates, 
denoted by $r=\{r_i\}$, $i=1,2,3$, are the coordinates of the particles 
in the current configuration. 
The two coordinates are related through the following equations
\begin{eqnarray}
r & = & r (a,t)\ ,\\
a & = & a (r,t)\ .
\end{eqnarray}
The velocity of the material element that is currently located at the 
point $r$ is defined as the time rate of change of its position
\begin{eqnarray}
v & = & \left( \frac{\partial r} {\partial t} \right)_a\ ,
\label{velocity-1}
\end{eqnarray} 
where the subscript $a$ is used to emphasize the fact that the 
derivative is to be evaluated for a particular material element
whose Lagrangian coordinates are $a$.
This is the usual material derivative.
In the Eulerian description, the material derivative of any property
pertaining to the particle labeled by the Lagrangian coordinate $a$ 
is given by
\begin{eqnarray}
\frac{D}{D t} & = & \frac{\partial}{\partial t} + v \cdot \nabla\ ,
\label{material-1}
\end{eqnarray}
where $v$ is the velocity of the particle $a$ at position $r$, given 
in Eq.~(\ref{velocity-1}).
For small deformations in which the Lagrangian strain tensor and
the Eulerian strain tensor have the same form,
the strain tensor $\epsilon$ is the symmetric part of the
displacement gradient tensor,
\begin{eqnarray}
\epsilon & = & \frac{1}{2} \left[ \nabla (\Delta r)
+ \nabla (\Delta r)^{\dagger} \right]\ ,
\label{strain-1}
\end{eqnarray}
where $\Delta r = r - a$, and the dagger superscript stands 
for the transposition of tensors.
The material derivative of the strain tensor is given
by~\cite{Reismann},
\begin{eqnarray}
\frac{D \epsilon}{D t} & = & \frac{1}{2} \left(
\nabla v + \nabla v^{\dagger}\right)\ .
\label{derivative-10}
\end{eqnarray}
Finally, the principle of conservation of mass leads directly to the
following well-known formula~\cite{Reismann,Mal,Fung}
\begin{eqnarray}
\frac{d}{d t} \int d^3 r \rho (r,t) Q (r,t) & = &
\int d^3 r \rho (r,t) \frac{D}{D t} Q (r,t)\ ,
\label{derivative}
\label{integral-0}
\end{eqnarray}
where $Q (r,t)$ is any physical quantity per unit mass and 
$\rho (r,t)$ is the mass density of the material.

We now generalize the above concepts to the two-fluid picture of
polymer blends.
As usual, we choose the undeformed state as the reference configuration,
in which each material particle is identified by its Lagrangian
coordinates $a$.
Since in the present situation there are two kinds of material
particles moving with different velocities in the system,
one should distinguish the Lagrangian coordinates for the two species.
We denote the Lagrangian coordinates of the particles of polymers $A$ 
and $B$ by $a_A$ and $a_B$ respectively.
Then Eq.~(\ref{velocity-1}) can be generalized as
\begin{eqnarray}
v_A & = & \left( \frac{\partial r} {\partial t} \right)_{a_A}\ ,
\label{velocity-21}\\
v_B & = & \left( \frac{\partial r} {\partial t} \right)_{a_B}\ .
\label{velocity-22}
\end{eqnarray}
The physical meaning of Eqs.~(\ref{velocity-21}) and~(\ref{velocity-22}) 
is as follows.
At each space-time point in the current configuration, 
there are two velocities, $v_A (r,t)$ and $v_B (r,t)$, which will 
be acquired by the material particles passing through this point, 
depending on the type of material particles.
That is, particles of polymer $A$ pass the point with velocity $v_A (r,t)$,
while particles of polymer $B$ pass the same point with velocity
$v_B (r,t)$.
The fluid velocity (average velocity) of this point is given by
\begin{eqnarray}
v (r,t) & = & \phi (r,t) v_A (r,t) + \left[1-
\phi (r,t)\right] v_B (r,t)\ .
\label{velocity-average}
\end{eqnarray}
Naturally, corresponding to Eqs.~(\ref{velocity-21}) 
and (\ref{velocity-22}), we may introduce two kinds of material 
derivatives in the system
\begin{eqnarray}
(\frac{D}{D t})_A & = & \frac{\partial}{\partial t} + v_A \cdot \nabla\ ,
\label{material-21}\\
(\frac{D}{D t})_B & = & \frac{\partial}{\partial t} + v_B \cdot \nabla\ .
\label{material-22}
\end{eqnarray}
If we focus on particles of polymer $A$, the material derivative is 
given by Eq.~(\ref{material-21}); 
similarly, the material derivative for species $B$ is given by 
Eq.~(\ref{material-22}).
The essential point is that
any difference between $v_A$, $v_B$, and the center of mass
velocity is due to mutual diffusion.
A constitutive relation is required to fix the 
diffusion flux or, equivalently, $v_A - v_B$.
We will return to this point later.

It is easy to evaluate the two material derivatives for some
basic physical quantities, such as the volume fraction $\phi$
and strain tensor $\epsilon$.
First, from the continuity equations for both species 
\begin{eqnarray}
\frac{\partial \rho_A (r,t)}{\partial t} 
+ \nabla \cdot \rho_A (r,t) v_A (r,t) & = & 0 ,
\label{con-11}\\
\frac{\partial \rho_B (r,t)}{\partial t} 
+ \nabla \cdot \rho_B (r,t) v_B (r,t) & = & 0 \ ,
\label{con-12}
\end{eqnarray}
we can obtain the expressions for the material derivatives of
the volume fraction
\begin{eqnarray}   
\left[\frac{D \phi(r,t)}{D t}\right]_{A} & = & - \phi (r,t)
\nabla \cdot v_A (r,t) \ ,
\label{con-31}\\   
\left[\frac{D \phi(r,t)}{D t}\right]_{B} & = & [1- \phi (r,t)]
\nabla \cdot v_B (r,t) \ ,
\label{con-32}
\end{eqnarray}
where Eq.~(\ref{assumption-1}) has been used.
Note that augmented by  Eq.~(\ref{assumption-1}), the continuity 
equations lead directly to the incompressibility condition
\begin{eqnarray}
\nabla \cdot v & = & \nabla \cdot \phi v_A 
+ \nabla \cdot (1-\phi) v_B\ =\ 0\ .
\label{incompressible}
\end{eqnarray}

Next, for small deformations, the strain tensor of the polymer
blend is still given by Eq.~(\ref{strain-1}), but
the material derivative of $\epsilon$ is generalized to the following
equations,
\begin{eqnarray}
\left(\frac{D \epsilon}{D t}\right)_{i} & = & \frac{1}{2} \left(
\nabla v_i + \nabla v_i^{\dagger}\right)\ ,\ i=A,B\ .
\label{derivative-1}
\end{eqnarray}
Taking material derivatives defined in Eqs.~(\ref{material-21}) 
and~(\ref{material-22}) on both sides of Eq.~(\ref{strain-1}), 
and noticing that for small deformation,
$\nabla = \partial /\partial r \simeq \partial /\partial a $,
Eq.~(\ref{derivative-1}) is obtained.
As will be seen in Subsection C, Eqs.~(\ref{con-31}), (\ref{con-32}), 
and (\ref{derivative-1})
are useful in the evaluation of the dissipation rate of the total
free energy of the system.

Finally, it is follows from the principle of mass conservation that
Eq.~(\ref{integral-0}) still holds.
But in the present situation, $\rho$ is the total mass density 
$\rho = \rho_A + \rho_B$, which is a constant 
(see Eq.~(\ref{assumption-1})), and  
$v$ is the fluid velocity given in Eq.~(\ref{velocity-average}).
Noticing the fact that the system is incompressible 
($\nabla \cdot v = 0$), for a polymer blend Eq.~(\ref{integral-0}) can 
be written as
\begin{eqnarray}
\frac{d}{d t} \int d^3 r \rho Q (r,t) & = &
\int d^3 r \rho \frac{\partial Q}{\partial t}\ .\
\label{integral-00}
\end{eqnarray}
Here, a boundary term has been ignored. 
Furthermore, since the masses of both species are also conserved 
individually, one has the following equations
\begin{eqnarray}
\frac{d}{d t} \int d^3 r \rho_{i} Q_i (r,t) & = &
\int d^3 r \rho_{i} \left(\frac{D}{D t}\right)_{i} Q_i (r,t)\ ,\ 
i=A,B\ ,
\label{integral-1}
\end{eqnarray}
where $\rho_i Q_i$ is any physical quantity contributed by the 
$i$-species.
Note that only two of the three formulae for the time derivative of 
the volume integration are independent.
In fact, summing up the two equations given in Eq.~(\ref{integral-1}),
one obtains Eq.~(\ref{integral-0}).

\vspace{0.3cm}

{\bf B. Total Free Energy}

\vspace{0.3cm}

We take into account three kinds of contributions to the total 
free energy of the system: The kinetic energy $K$ of moving particles,
the mixing free energy $F_m$ of the two species, and the elastic free 
energy $F_e$ of polymers due to chain deformation~\cite{Onuki}.
Thus, the total free energy of the system can be written as 
\begin{eqnarray}
F_t & = & K\ +\ F_m (\phi) \ +\ F_e (\phi, \epsilon)\ .
\label{f-1}
\end{eqnarray}
Here, we have assumed that the mixing free energy is a function
of $\phi$ only, while the elastic free energy depends on both  
$\phi$ and $\epsilon$.
The kinetic energy of the two kinds of moving particles can be 
expressed as
\begin{eqnarray}
K & = & \int d^3 r\ \left(\frac{1}{2} \rho_A  v_A^2
+\ \frac{1}{2} \rho_B  v_B^2\right)\ .
\label{f-1k}
\end{eqnarray}
The mixing free energy can be written as
\begin{eqnarray}
F_m (\phi) & = &  \int d^3 r f_m (\phi)\ ,
\label{free-0bm}
\end{eqnarray}
where $f_m (\phi)$ is the mixing free energy density, which,
for example, can be chosen
to be the Flory-Huggins form.
Our derivation for the kinetic equations is independent of the 
precise form of $f_m (\phi)$.  
In a similar way, the elastic free energy can be expressed as 
\begin{eqnarray}
F_e (\phi,\epsilon) & = &  \int d^3 r f_e (\phi,\epsilon)\ .
\label{free-0b}
\end{eqnarray}
In the theory of linear elasticity, the elastic energy density due to 
chain deformation can be phenomenologically expressed 
as~\cite{Onuki,taoo},
\begin{eqnarray}
f_{e}(\phi,\epsilon) & = & G (\phi)\ f_e^{*} (\epsilon)\ ,
\label{fe-0}
\end{eqnarray}
where $f_e^{*}(\epsilon) = \epsilon:\epsilon$, and the coefficient $G$ 
is the shear modulus of the system, which, in general, depends
on concentration only.
Here, the notation $(:)$ stands for the scalar product of second
rank tensors.

Since in a polymer blend both species contribute to the elastic energy, 
we propose the following intuitive form for the shear modulus of the
blend
\begin{eqnarray}
G  & = &  G_B^{(0)} +\  
\left[ G_A^{(0)} - G_B^{(0)} \right] \ 
\Delta (\phi)\ ,
\label{gb0}
\end{eqnarray}
where $G_i^{(0)}$ is the shear modulus of the $i$-species before 
mixing (``bare'' shear modulus), and 
$\Delta (\phi)$ is an interpolating function describing the 
effect of blending. 
The condition that $G(\phi = 0) = G_B^{(0)}$ and $G(\phi = 1) = G_A^{(0)}$
requires that $\Delta (0) = 0$ and $\Delta (1) = 1$.
Then the simplest form of the interpolating function would be
$\Delta(\phi) = \phi$, i.e., an ideal mixture approximation.
More generally we suppose
\begin{eqnarray}
\Delta (\phi) & = &  \phi\ + \tilde{\Delta} (\phi)\ 
\label{Delta}
\end{eqnarray}
with $\tilde{\Delta} (0) = \tilde{\Delta} (1) = 0$.
For use in Eq.~(\ref{fe-0}), it is convenient to cast 
Eq.~(\ref{gb0}) into the form
\begin{eqnarray}
G (\phi) & = &  \phi G_A (\phi)\ +\ (1-\phi) \ G_B (\phi)\ ,
\label{gb}
\end{eqnarray}
where $G_i$, $i=A,B$, are the ``renormalized" shear moduli of 
species $A$ and $B$ individually, which can be expressed as
\begin{eqnarray}
G_A (\phi) & = &  G_A^{(0)} \left[1 + \frac{\tilde{\Delta} (\phi)}
{\phi} \right]\ ,
\label{GA}\\
G_B (\phi) & = &  G_B^{(0)} \left[1 - \frac{\tilde{\Delta} (\phi)}
{1-\phi} \right]\ .
\label{GB}
\end{eqnarray}
As we will see below, Eq.~(\ref{gb}) is a reasonable approximation
leading to a sensible result for the network velocity.

Finally, since the polymer blend is viscoelastic (rather than a purely
elastic system), one expects that Eq.~(\ref{f-1}) describes the
physics within the time scales less than the relaxation time
of the shear stress (usually referred to as the ``terminal relaxation
time")~\cite{Gennes}. 

\vspace{0.3cm}
 
{\bf C. Dissipation Rate of Total Free Energy}
 
\vspace{0.3cm}

With the results presented in the previous two subsections, we
are ready to discuss the dissipation rate of the total free energy
of the system.
First, using the formula given in Eq.~(\ref{integral-1}), the evaluation
of the time derivative of the kinetic energy is straightforward, and
the result can be expressed as
\begin{eqnarray}
\frac{d K}{d t} & = & \int d^3 r\ \left\{\rho_A  v_A
\cdot \left(\frac{D v_A }{D t}\right)_{A} + \rho_B  v_B
\cdot \left(\frac{D v_B }{D t}\right)_{B}
\right\}\ .
\label{fd-k}
\end{eqnarray}
Next, since, in general, the mixing free energy $f_m(\phi)$ cannot be 
simply divided into contributions by particles $A$ and $B$, it is convenient
to use Eq.~(\ref{integral-00}) to calculate the dissipation rate of 
the mixing free energy.
Indeed, it is easy to obtain
\begin{eqnarray}
\frac{d F_m}{d t} & = & \int d^3 r\
\frac{\partial f_m (\phi) }{\partial t} \ = \ 
\int d^3 r\ \frac{\partial f_m (\phi) }{\partial \phi}\
\frac{\partial \phi}{\partial t} \ .
\label{fd-m}
\end{eqnarray}
Making use of the continuity equation (\ref{con-31}) and integrating by 
parts, we have
\begin{eqnarray} 
\frac{d F_m}{d t} & = & \int d^3 r\  v_A \cdot \phi \nabla
\frac{\partial f_m (\phi) }{\partial \phi} \ = \  
\int d^3 r\ v_A \cdot \nabla \pi_m\ , 
\label{fd-m1} 
\end{eqnarray} 
where $\pi_m$ is the osmotic pressure associated with the mixing free 
energy, given by
\begin{eqnarray}
\pi_{m} & = & \left(\phi \frac{\partial}{\partial \phi}-1
\right) f_{m} (\phi)\ .
\label{pi-m}
\end{eqnarray}
Note that if the continuity equation (\ref{con-32}) were used, 
a different expression for $d F_m/d t $ would be obtained, 
but it will give the same final kinetic equations when the 
condition $\nabla \cdot v = 0$ is taken into account.

Finally, we discuss the time derivative of the elastic free energy.
In view of Eq.~(\ref{gb}), we can write
\begin{eqnarray}
f_e (\phi,\epsilon) & = &  f_{eA} (\phi,\epsilon) 
+ f_{eB} (\phi,\epsilon)\ ,
\label{free-0c}
\end{eqnarray}
where $f_{ei} (\phi,\epsilon) = \phi_i G_i f^{*}_e (\epsilon)$, for 
$i=A,B$, are the elastic free energy densities contributed
by the two components individually.  
Using Eq.~(\ref{integral-1}), we have
\begin{eqnarray}
\frac{d F_e}{d t} & = & \int d^3 r\ \left\{
\phi \left[\frac{D}{D t} \phi^{-1} 
f_{eA} (\phi,\epsilon)\right]_A 
+ (1-\phi) \left[\frac{D}{D t} (1-\phi)^{-1} 
f_{eB} (\phi,\epsilon) \right]_B \right\}\ .
\label{fd-e1}
\end{eqnarray}
Using the chain rule,
the material derivatives of the free energies $f_{eA} (\phi,\epsilon)$ and 
$f_{eB} (\phi,\epsilon)$ can be expressed 
in terms of the material derivatives of $\phi$ and $\epsilon$.
Making use of Eqs.~(\ref{con-31}), (\ref{con-32}), 
and~(\ref{derivative-1}), we obtain
\begin{eqnarray}
\frac{d F_e}{d t} & = & \int d^3 r \left\{ -\pi_{eA} \nabla \cdot v_A
-\pi_{eB} \nabla \cdot v_B 
+ (\nabla v_A):\frac{\partial f_{eA}}{\partial \epsilon} 
+ (\nabla v_B):\frac{\partial f_{eB}}{\partial \epsilon}
\right\}\ ,
\label{rate-e2}
\end{eqnarray}
where $\pi_{ei}$, with $i = A, B$, are the ``elastic osmotic pressures",  
given by
\begin{eqnarray}
\pi_{ei} & = & \left(\phi_i \frac{\partial}{\partial \phi_i}-1
\right) f_{ei} (\phi,\epsilon)\ ,\ i=A,B\ .
\label{pi-i}
\end{eqnarray}
The stress tensor acting on the network can be defined as~\cite{Landau}
\begin{eqnarray}
\tau & = & \frac{\partial f_{e} (\phi,\epsilon)}
{\partial \epsilon}\ .
\label{tau-i}
\end{eqnarray}
In view of Eqs.~(\ref{fe-0}) and~(\ref{free-0c}), 
it is easy to check that
\begin{eqnarray}
\frac{\partial f_{eA}}{\partial \epsilon}\ 
& = & \phi \frac{G_A}{G} \ \tau\ ,
\label{pe-1}\\
\frac{\partial f_{eB}}{\partial \epsilon}\  
& = & (1-\phi) \frac{G_B}{G} \ \tau\ . 
\label{pe-2}
\end{eqnarray}
Substituting Eqs.~(\ref{pe-1}) and~(\ref{pe-2}) into 
Eq.~(\ref{rate-e2}), the dissipation rate of the elastic free energy
can be written as
\begin{eqnarray}
\frac{d F_e}{d t} & = & \int d^3 r \left\{ 
v_A \cdot \left[ \nabla \pi_{eA} -
\nabla \cdot \phi \frac{G_A}{G} \tau \right]+
v_B \cdot \left[ \nabla \pi_{eA} - 
\nabla \cdot (1-\phi) \frac{G_A}{G} \tau \right]
\right\}\ , 
\label{rate-e3}
\end{eqnarray}
after an integration by parts. 

Combining Eqs.~(\ref{fd-k}), (\ref{fd-m1}), and (\ref{rate-e3}),
we obtain the dissipation rate of the total free energy
\begin{eqnarray}
\frac{d F_t}{d t} & = & \int d^3 r \left\{\rho_A v_A \cdot
\left(\frac{D v_A}{D t}\right)_{A} + \rho_B
v_B \cdot \left(\frac{D v_B}{D t}\right)_{B}
\right.\nonumber\\ 
& + & \left. 
v_A \cdot \left[\nabla (\pi_m+ \pi_{eA})
-\nabla \cdot \phi \frac{G_A}{G} \tau \right]\right.\nonumber\\
& + & \left.
v_B \cdot \left[\nabla \pi_{eB}
-\nabla \cdot (1-\phi) \frac{G_B}{G} \tau \right] \right\}\ .
\label{rate-final}
\end{eqnarray}
It should be understood that the partial derivative  
$\partial/\partial \phi$ is carried out at fixed $\epsilon$
and $\partial/\partial \epsilon$ is carried out at fixed $\phi$.
 
\vspace{0.3cm}
 
{\bf D. Network Velocity}
 
\vspace{0.3cm}

In this subsection, we discuss the network velocity (or tube velocity
in the reptation picture).
Substituting Eqs.~(\ref{pe-1}) and (\ref{pe-2}) into Eq.~(\ref{rate-e2}),
the dissipation rate for the elastic free energy 
can be expressed as
\begin{eqnarray}
\frac{d F_e}{d t} & = & \int d^3 r \left\{
-\pi_{eA} \nabla \cdot v_A -\pi_{eB} \nabla \cdot v_B 
\right. \nonumber\\
& + & \left. G^{-1} \left[ \phi G_A \nabla v_A
+ (1-\phi) G_B \nabla v_B \right]:\tau \right\}\ .
\label{rate-f}
\end{eqnarray}
The last term in the above equation describes the time rate of change
of the elastic energy purely due to the change of the strain tensor, 
so that the coefficient of the stress tensor $\tau$ can be identified as the
gradient of the network velocity (or tube velocity in the reptation
picture),
\begin{eqnarray}
\nabla v_t & = &  G^{-1} \left[ \phi G_A \nabla v_A
+ (1-\phi) G_B \nabla v_B \right]\ .
\label{vt-1}
\end{eqnarray}
As we now show, this expression is in agreement with a previously
obtained result using a microscopic approach.

The network velocity $v_t$ has previously been estimated from molecular 
theory ~\cite{Doi1,Brochard} for a uniform system 
($\phi$ is constant) with the result
\begin{eqnarray}
v_t & = & \frac{\zeta_A v_A + \zeta_B v_B}{\zeta_A +\zeta_B}\ .
\label{vt-0}
\end{eqnarray}
Here $\zeta_i$, $i=A,B$, are effective friction coefficients
given by 
\begin{eqnarray}
\zeta_i & = & \nu_i \phi_i \frac{N_i}{N_i^{e}} \zeta_0\ ,\ i=A,B\ ,
\label{zetai}
\end{eqnarray}
where $\nu_i$ is the number of chains of species $i$ per unit volume,
$N_{i}^{e}$ is the average interval between two successive 
entanglement points along one chain, and $\zeta_0$ is the 
phenomenological friction coefficient between the two species.
On the other hand, when $\phi$ is independent of space, 
Eq.~(\ref{vt-1}) can be solved with the result   
\begin{eqnarray} 
v_t & = &  G^{-1} \left[\phi G_A v_A
+ (1-\phi) G_B v_B \right]\ ,
\label{vt-1a}
\end{eqnarray}
where an integrating constant has been determined as zero
from the condition that $v_t = v_A$, when $\phi=1$.
We may suppose that $G_{A,B}$ measure the densities of entanglement
points, i.e., $G_i \propto T \phi_i / N_i^e$~\cite{Gennes}.
Then Eq.~(\ref{vt-0}) is recovered.

\vspace{0.3cm}

{\bf E. Kinetic Equations}

\vspace{0.3cm}

First we derive the equations for the two velocity fields 
$v_A$ and $v_B$.
These equations can be obtained by means of 
Rayleigh's variational principle~\cite{Doi0,Doi1}.
Following Doi and Onuki~\cite{Doi0,Doi1}, one can define a
Rayleighian functional,  
\begin{eqnarray}
R & = & \frac{1}{2}\ W \ +\ \frac{d F_t}{d t}\ ,
\label{functional}
\end{eqnarray}
where $F_t$ is the total free energy of the system and $W$ is the 
dissipation function due to relative motion of the two polymers, 
which one assumes can be written as
\begin{eqnarray}
W & = & \int d^3 r\ c(r) \zeta\  (v_A - v_B )^2\ .
\label{dissipation}
\end{eqnarray}
Here $c$ is the monomer concentration of species $A$ defined 
via $\phi=v_m c$ with $v_m$ the monomer volume, and $\zeta$ is the 
friction constant, which, in general, is a function of $\phi$~\cite{Milner}. 
The Rayleighian defined in Eq.~(\ref{functional}) may be understood
as the total energy dissipation rate of the system. 
The variational principle states that the velocities 
$v_A$ and $v_B$ are determined by the condition that they minimize
the Rayleighian~\cite{Doi0,Doi1}. 

Substituting Eqs.~(\ref{rate-final}) and~(\ref{dissipation}) 
into Eq.~(\ref{functional}), we have the Rayleighian for the 
polymer blend
\begin{eqnarray}
R & = & \int d^3 r \left\{
\rho_A v_A \cdot \left(\frac{D v_A}{D t}\right)_A +
\rho_B v_B \cdot \left(\frac{D v_B}{D t}\right)_B \right.\nonumber\\
& + & \left. v_A \cdot \left[\nabla (\pi_m+ \pi_{eA})
-\nabla \cdot \phi \frac{G_A}{G} \tau \right]\right.\nonumber\\
& + & \left. v_B \cdot \left[\nabla \pi_{eB}
-\nabla \cdot (1-\phi) \frac{G_B}{G} \tau \right]\right.\nonumber\\
& + & \left. \frac{1}{2} c (r) \zeta\ (v_A - v_B)^2 \right\}\ .
\label{functional-1b}
\end{eqnarray}
Since $v_A$ and $v_B$ are not independent variables due to the
incompressibility condition~(\ref{incompressible}), the functional 
$R$ must be minimized under this condition. 
The conditional minimization of the functional $R$ with 
respect to $v_A$ and $v_B$ leads to the following equations,
\begin{eqnarray}
\rho_A \left(\frac{D v_A}{D t}\right)_A & = &
- c \zeta (v_A - v_B) - (\nabla p) \phi - \nabla (\pi_m + \pi_{eA})
+ \nabla \cdot \phi \frac{G_A}{G} \tau \ ,
\label{ns-lb}\\
\rho_B \left(\frac{D v_B}{D t}\right)_B & = &
c \zeta (v_A - v_B) - (1-\phi) (\nabla p)  - \nabla \pi_{eB}\ 
+ \nabla \cdot (1-\phi) \frac{G_B}{G} \tau \ ,
\label{ns-sb}
\end{eqnarray}
where $p$ is the Lagrange multiplier imposing the incompressibility
condition.
Eqs.~(\ref{ns-lb}) and~(\ref{ns-sb}) describe the motion of polymers
$A$ and $B$ in the system.

Solving Eqs.~(\ref{ns-lb}) and~(\ref{ns-sb}) for $v_A$ and then 
substituting the resulting expression into Eq.~(\ref{con-31}), we can
obtain the diffusion equation for $\phi$.
Clearly this cannot be done exactly and some approximation must be
applied~\cite{Doi1}. 
Since the velocities relax much faster than $\phi$, we can ignore
the inertia terms in Eqs.~(\ref{ns-lb}) and~(\ref{ns-sb}) to obtain
an explicit expression for $v_A$.
After eliminating the $p$ terms, we can express $v_A$ 
as
\begin{eqnarray}
v_A & = & v - \frac{\phi(1-\phi)^2}{c \zeta} \left\{
\nabla \frac{\partial f}{\partial \phi} - \alpha \left[
(\nabla \epsilon):\tau + \nabla \cdot \tau \right] 
- \beta \cdot \tau \right\}\ ,
\label{vl-1}
\end{eqnarray}
where $f=f_m+f_e$ and Eq.~(\ref{velocity-average}) has been used.
In Eq.~(\ref{vl-1}), we have introduced two parameters $\alpha (\phi)$
and $\beta (\phi)$ for convenience, which are given by
\begin{eqnarray}
\alpha & = & G^{-1} \left[G_A - G_B \right]\ ,
\label{alpha}\\
\beta & = & \phi^{-1} \nabla \phi \frac{G_A}{G}
- (1-\phi)^{-1} \nabla (1-\phi) \frac{G_B}{G}\ .
\label{beta}
\end{eqnarray}
Substituting Eq.~(\ref{vl-1}) into Eq.~(\ref{con-31}), we obtain
\begin{eqnarray}
\frac{\partial \phi}{\partial t} + v \cdot \nabla \phi & = &
v_m \nabla \cdot \frac{\phi (1-\phi)^2}{\zeta} \left\{ 
\nabla \frac{\partial f}{\partial \phi} - \alpha \left[
(\nabla \epsilon):\tau + \nabla \cdot \tau \right] 
- \beta \cdot \tau \right\}\ ,
\label{eq1a}
\end{eqnarray}
where the incompressibility condition has been used.
This is the diffusion equation for $\phi$.

It is convenient to describe the motion of the system using the
fluid velocity $v$ and relative velocity $u = v_A -v_B$. 
From Eqs.~(\ref{ns-lb}) and~(\ref{ns-sb}), it is easy to check that
the kinetic equations for $v$ and $u$ can be written respectively as
\begin{eqnarray}
\rho \left(\frac{\partial v}{\partial t}
+  v \cdot \nabla v + R_{v}\right) 
& = & - \nabla p - \nabla \pi + \nabla \cdot \tau\ ,
\label{ns-vb}\\
\rho \phi \left(\frac{\partial u}{\partial t} 
+ u \cdot \nabla u + R_{u}\right)
& = & - \frac{c \zeta}{1-\phi}u -\phi
\left\{\nabla \frac{\partial f}{\partial \phi} - \alpha \left[
(\nabla \epsilon):\tau + \nabla \cdot \tau \right] 
- \beta \cdot \tau 
\right\}\ ,
\label{ns-ub}
\end{eqnarray}
where $\pi=\pi_m+\pi_{eA}+\pi_{eB}$ is the total osmotic pressure, 
and $R_{v}$ and $R_{u}$ are couplings between $v$ and $u$ that can be
expressed as 
\begin{eqnarray}
R_v & = & u \nabla \cdot \phi (1-\phi) u
+ \phi (1-\phi) u \cdot \nabla u\ , \\
R_u & = & u\cdot \nabla (v-\phi u) + (v-\phi u) \cdot \nabla u\ .
\end{eqnarray}
Eq.~(\ref{ns-vb}) is the generalized Navier-Stokes equation for the 
fluid velocity.
$R_v$ is the correction due to coupling between bulk flow and
relative motion.
Eq.~(\ref{ns-ub}) describes the relative motion between the two species.

The constitutive equation cannot be derived from the above formalism,
because so far no ordinary viscosity effects have been included
in the Rayleighian.
We assume that the time evolution of the stress tensor is described by
the upper convected Maxwell equation~\cite{Doi1,tao,Larson}
\begin{eqnarray}
\lambda \left[\frac{\partial \sigma}{\partial t}+v_t \cdot \nabla \sigma  
- \sigma \cdot \nabla v_t -
(\nabla v_t)^{\dagger} \cdot \sigma \right]
+\sigma & = & G (\phi) \delta\ ,
\label{eq3}
\end{eqnarray}
where $\lambda$ is the relaxation time, $v_t$ is the network velocity,
and the stress tensor $\sigma$ is related to $\tau$ through 
$\sigma = G \delta + \tau$.
The network velocity $v_t$ has been used here 
because the stress acts on the polymer network~\cite{Doi1,Doi}.
Since in polymer blends the fluctuations of concentration are
small, and Eq.~(\ref{vt-1}) cannot be solved explicitly,
we will use Eq.~(\ref{vt-1a}) as an approximate expression for
the network velocity $v_t$ in the linear analysis.
Using the expressions for $v_A$ and $v_B$, we have finally
\begin{eqnarray}
v_t & = & v - v_m \alpha \frac{\phi(1-\phi)^2}{\zeta}
\left\{\nabla \frac{\partial f}{\partial \phi} - \alpha \left[
(\nabla \epsilon):\tau + \nabla \cdot \tau \right]
- \beta \cdot \tau \right\}\ .
\label{vt-f}
\end{eqnarray}
Eqs.~(\ref{eq1a}), (\ref{ns-vb}), (\ref{ns-ub}), and~(\ref{eq3}) 
describe the dynamics of phase behavior of polymer blend under shear flow.
Since $u$ appears only in Eqs.~(\ref{ns-vb}) and (\ref{ns-ub}), and 
in most cases $R_v$ is negligibly small~\cite{Doi1}, one can ignore 
the $R_v$ term in Eq.~(\ref{ns-vb}), and then the model reduces to 
three equations.
In the case where the $B$-component represents small molecules,
$G_B \sim 0$, $\alpha \simeq 1/\phi$, $\beta \simeq 0$, 
the model for a polymer solution is recovered~\cite{Onuki,tao}. 

Note that although the structure of our set of kinetic equations is similar
to that given in Ref.~\cite{Doi1}, some important differences exist.
First, the free energy density $f$ in Eq.~(\ref{eq1a}) is the total free
energy density, i.e., $f = f_m + f_e$, while in Eq. (4.17) of 
Ref.~\cite{Doi1}, $f$ is the mixing free energy only.
That is, we have taken into account the fact that, in general, the 
elastic free energy can be $\phi$ dependent (through the shear 
modulus $G(\phi$)).
Second, two more terms ($\alpha (\nabla \epsilon):\tau)$ and 
($\beta \cdot \tau$) are generated in Eq.~(\ref{eq1a}).
The $\alpha (\nabla \epsilon):\tau)$ term and the inclusion 
of $f_e$ in the total free energy density $f$ ensure the 
that kinetic equations to reduce appropriately to the solution case.

\vspace{0.5cm}
 
{\bf III. LINEAR ANALYSIS} 

\vspace{0.5cm}
 
Eqs.~(\ref{eq1a}), (\ref{ns-vb}), and~(\ref{eq3}) are nonlinear 
equations, so that a complete analysis is difficult.
However, to study the effect of viscoelasticity on the phase boundary
under shear, it is sufficient to carry out a linear analysis similar 
to that developed in Refs.~\cite{Helfand,Onuki}.
In this formalism, we first iteratively solve Eq.~(\ref{eq3}) for 
$\sigma$ to the ``second-order fluid'' level
from which the stress tensor $\sigma$ 
can be expressed in terms of $\phi$ and $v$~\cite{Larson}.
Then we substitute this constitutive relation for $\sigma (v,\phi)$ 
into Eqs.~(\ref{eq1a}) and~(\ref{ns-vb}).
Setting 
\begin{eqnarray}
\phi & = & \phi_0 + \phi_1\ , 
\label{linear-phi0}\\ 
v    & = & v_0 + v_1\ , 
\label{linear}
\end{eqnarray}
where $\phi_0$ and $v_0$ are the overall average volume fraction and
the fluid velocity, and $\phi_1$ and $v_1$ are small deviations, 
we can solve the equations to linear order in $\phi_1$ and $v_1$.
In the case of shear flow, $v_0 = S y e_x$, where $S$ is the 
shear rate and $e_x$ is the unit vector in the $x$-direction.
The 
expansion parameter is essentially the combination $S\lambda$,
which limits the approximations to the regime of weak shear.

\vspace{0.3cm}

{\bf A. Iterations for $\sigma$ and $\epsilon$}

\vspace{0.3cm}

We first solve the Maxwell equation iteratively~\cite{Larson}.
Since our aim is to see how viscoelasticity changes the 
stability line for phase separation, it is sufficient to obtain a solution
for $\sigma$ in the long wavelength limit.
It is easy to check by ``power counting'' 
that $\phi_1$ and $\nabla v_1$
are the leading order terms, and then 
$v -v_{t} \sim \nabla \phi + \nabla^2 v$ 
are higher order terms.
Therefore, we can replace $v_t$ by $v$ in the Maxwell equation for
the present purpose.
Furthermore, since $v$ and $\sigma$ relax much faster than $\phi$,
we may ignore the inertia terms~\cite{Helfand,Onuki} and obtain 
\begin{eqnarray} 
\lambda \left[- \sigma \cdot \nabla v -
(\nabla v)^{\dagger} \cdot \sigma \right]
+\sigma & = & G (\phi) \delta\ .
\label{eq3a}  
\end{eqnarray}
Directly iterating 
to the ``second-order fluid'' (i.e., to ${\cal O} (|\nabla v|^2)$, 
we obtain
\begin{eqnarray} 
\sigma & = & G \delta +\ 2 \eta D 
+ 2 \eta \lambda \left(D \cdot \nabla v 
+ \nabla v^{\dagger} \cdot D \right)\ , 
\label{sigma-1}
\end{eqnarray} 
where $\eta = \lambda G$ is the viscosity and $D$ is the gradient
tensor given by $2 D = \nabla v + \nabla v^{\dagger}$. 
Note that, at this order, only leading order non-Newtonian
terms are included.
Finally, for later convenience, we write down the expression for
the strain tensor $\epsilon$
\begin{eqnarray}  
\epsilon & = &  \lambda D 
+ \ \lambda^2 \left(D \cdot \nabla v 
+ \nabla v^{\dagger} \cdot D \right)\ .
\label{epsilon-1}  
\end{eqnarray}   
Now we have expressed $\sigma$ in terms of $\phi$ and $v$.
Our next task is to eliminate $\sigma$ and $\epsilon$ using 
Eqs.~(\ref{sigma-1}) and~(\ref{epsilon-1}) in Eqs.~(\ref{eq1a})
and~(\ref{ns-vb}) and carry out a linear analysis for the remaining
equations. 

\vspace{0.3cm}

{\bf B. The Navier-Stokes Equation}

\vspace{0.3cm}

Now we discuss the linear analysis of the Navier-Stokes equation.
In a similar spirit to the treatment of the Maxwell equation, we 
ignore the inertia terms since fluid velocities
relax much faster than $\phi$~\cite{Helfand,Onuki}.
Then the equation for $v$ becomes simply
\begin{eqnarray}
\nabla p + \nabla \pi - \nabla \cdot \tau & = & 0\ .
\label{nv-00}
\end{eqnarray}
We next find expressions for $\nabla \pi$ and $\nabla \cdot \tau$
linear in $\phi_1$ and $v_1$.
For later convenience, we introduce two parameters $g$ and $\xi$
through $G (\phi) = k_B T g (\phi)$ and
$\eta (\phi) = k_B T \xi (\phi)$.
It follows from the definition that 
$
\pi=\pi (\phi,\epsilon) = \tilde{\pi} (\phi,v)
$, and therefore we have 
\begin{eqnarray} 
\nabla \pi & = & 
\left(\frac{\partial \tilde{\pi}}{\partial \phi}\right)_0  \nabla \phi_1
+ (\nabla v_1) \cdot \left(\frac{\partial \tilde{\pi}}
{\partial v}\right)_0\ ,
\label{dot-pi1}
\end{eqnarray} 
where the subscript ``0'' indicates that the derivatives are evaluated
at $\phi_0$ and $v_0$. 
In view of Eq.~(\ref{sigma-1}), we can express 
$\nabla \cdot \tau$ in a similar way as
\begin{eqnarray}
\nabla \cdot \tau & = & k_B T \nabla \cdot 
\left[ 2 \xi D + 2 \xi \lambda \left(D \cdot \nabla v 
+ \nabla v^{\dagger} \cdot D \right) \right]\ .
\label{dot-tau}
\end{eqnarray}
Substituting Eqs.~(\ref{linear-phi0}) and~(\ref{linear}) into 
Eq.~(\ref{dot-tau}) and 
keeping terms only linear in $\phi_1$ and $v_1$, we have
\begin{eqnarray}  
\nabla \cdot \tau & = & k_B T \left[
\xi_0 \nabla^2 v_1 + S \xi_0' \left(e_x \partial_y \phi_1
+e_y \partial_x \phi_1\right) + {\cal O} (S^2,Sv) \right]\ , 
\label{dot-tau1}
\end{eqnarray}    
where $e_i$ with $i=x,y,z$ are unit vectors.
Here the subscript ``0'' means that the values are evaluated
at $\phi = \phi_0$, and primes indicate $\phi$-differentiation. 
To obtain the shift in the spinodal (stability line) to leading 
order in the shear rate 
(${\cal O} (S\lambda)^2$), the ${\cal O} (S^2,Sv)$ terms can be ignored
at this stage.   
 
Substituting Eqs.~(\ref{dot-pi1}) and (\ref{dot-tau1}) into 
Eq.~(\ref{nv-00}) and eliminating $p$ via 
$\nabla \cdot v = 0$, we can solve $v$ to order ${\cal O} (S)$ 
in Fourier space with the result 
\begin{eqnarray}
v_{1x} (k) & = & - i S \frac{\xi_0'}{\xi_0} \frac{k_y}{k^2}
\left(2 \hat{k}_x^{2} -1\right) \phi_1 (k) \ ,
\label{v1x}\\
v_{1y} (k) & = & - i S \frac{\xi_0'}{\xi_0} \frac{k_x}{k^2}
\left(2 \hat{k}_y^{2} -1\right) \phi_1 (k) \ ,
\label{v1y}\\
v_{1z} (k) & = & - 2 i S \frac{\xi_0'}{\xi_0} \frac{k_x k_y k_z}{k^4}
\phi_1 (k) \ ,
\label{v1z}
\end{eqnarray}
where $\hat{k}_i = k_i/k$.
As we see, to this order, the solution for $v$ is 
independent of the precise form of $\pi$.
Note that the expressions given in Eqs.~(\ref{v1x})-(\ref{v1z})
are the same as those for polymer solutions~\cite{Onuki,tao}.
That is, to leading order, the linear $\phi-v$ relations are 
the same for both polymer solutions and blends.

\vspace{0.3cm}

{\bf C. Shift in the Stability Line}

\vspace{0.3cm}

We now discuss the linearization of the diffusion equation~(\ref{eq1a}), 
from which the effect of viscoelasticity on the phase boundary can be studied.
We choose the Flory-Huggins form for the mixing free
energy~\cite{Gennes},
\begin{eqnarray}
f_{m} (\phi) & = & k_B T \left[\frac{\phi}{N_A} \ln \phi
+ \frac{1-\phi}{N_B} \ln (1-\phi) + \chi \phi (1-\phi) \right]\ ,
\label{fm-0}
\end{eqnarray}
where $\chi$ is the Flory-Huggins interaction parameter,
and use Eq.~(\ref{fe-0}) for the elastic free energy.
Substituting Eqs.~(\ref{linear-phi0}) and~(\ref{linear})
into Eq.~(\ref{eq1a}) and making use of Eqs.~(\ref{v1x})-(\ref{v1z}),
we can obtain the following linearized diffusion equation in Fourier space
\begin{eqnarray}
\left(\frac{\partial}{\partial t}
- S k_x \frac{\partial}{\partial k_y}\right) \phi_{1} (k)
& = & - v_m \frac{k_B T}{\zeta} \phi_0 (1-\phi_0)^2 k^2
\left[ \Psi_0 + \Psi_1 (\hat{k}_x,\hat{k}_y) \right] \phi_1 (k)\ .  
\label{linear-phif}
\end{eqnarray}
Here $\Psi_0$ is an isotropic constant given by
\begin{eqnarray}
\Psi_0 (\phi_0) & = &
\frac{1}{N_A \phi_0} + \frac{1}{N_B (1-\phi_0)}
-2 \chi + \kappa (\phi_0)  \left(S \lambda_0\right)^2\ ,
\end{eqnarray}
where $\lambda_0 = \xi_0/g_0$ is the average relaxation time and the
coefficient $\kappa$ is given by
\begin{eqnarray}
\kappa (\phi_0) & = & \frac{1}{2} g''_0
- \frac{g^{'2}_0}{g_0} + \alpha_0 g'_0\ .
\label{omega}
\end{eqnarray}
Note that $\kappa$ depends only on the shear modulus of the blend. 
The second term on the right-hand-side of Eq.~(\ref{linear-phif})
produces an anisotropic modification of the scattering 
function~\cite{Helfand} and is given by
\begin{eqnarray}  
\Psi_1 & = & - 2 S \lambda_0 (A_0 + B_0) \hat{k}_x \hat{k}_y
+ (S\lambda)^2 \left[ 4 (C_0 - 2 B_0) (\hat{k}_x \hat{k}_y)^2
\right.\nonumber\\
& - & \left. 2 ( A_0 + B_0 -D_0) \hat{k}_x^2 +
(C_0 - B_0) \hat{k}_z^2 \right]\ , 
\label{psi1}
\end{eqnarray}
where $A_0$, $B_0$, $C_0$, and $D_0$ are the constants
\begin{eqnarray}
A_0 & = &  \alpha_0' g_0 + \phi_0^{-1} g_{A0} + (1-\phi_0)^{-1} g_{B0}\ ,
\label{a0}\\
B_0 & = &  \alpha_0 g_0 \frac{\xi_0'}{\xi_0}\ ,
\label{b0}\\
C_0 & = & g_0' \frac{\xi_0'}{\xi_0}\ ,
\label{c0}\\
D_0 & = & \alpha_0 g_0' .
\label{d0}
\end{eqnarray}
The zero-shear spinodal line (linear stability 
line) in the $\chi-\phi$ 
parameter space is determined by the first three terms of 
$\Psi_0$~\cite{Gennes}.
The last term is the modification of the stability line arising from the 
dynamics of viscoelasticity; 
this shift is ${\cal O} (S^2)$, where $S$ is the
shear rate.

The direction of the stability line shift is dependent on the
sign of $\kappa$, which in turn depends on the details of
the interpolating function $\Delta(\phi)$.
When $\kappa>0$, the effective value of $\chi$ is reduced,
and stability line in the $\chi-\phi$ plane
is shifted to larger values of $\chi$ (lower temperatures).
On the other hand, when $\kappa<0$, the stability line 
is shifted in the opposite manner.
Since, in general, 
the shear moduli for the individual species $A$ and $B$ involve
material parameters, the details of the shift in the 
stability line cannot be determined explicitly. 
In highly symmetric situations, the magnitude of any shift might
be small, owing to the small value of the coefficient $[G_A^{(0)}-G_B^{(0)}]$.
However,
the shift in the linear stability should, according to the
present analysis, correlate with measurements of the 
interpolating function for the blend, $\Delta(\phi)$.
If $\Delta(\phi) = \phi$, $\kappa = 0$ and there is no shift in
stability.

\vspace{0.5cm}

{\bf IV. CONCLUSION}

\vspace{0.5cm}

We have applied the general two-fluid approach of Doi and Onuki~\cite{Doi1}
to establish model equations for the study of the phase behavior,
or more precisely, linear stability, of a polymer blend 
in the presence of shear flow.
A phenomenological form of the shear modulus for a polymer blend is
used, which can be experimentally determined. The modifications 
used here allow the kinetic equations to reduce to the solution case
in the limit in which one of the species becomes a small (solvent) molecule.
Linear analysis of the model indicates that the equilibrium stability 
line is shifted by the effect of 
viscoelasticity when the dependence of the shear modulus of the blend
on the volume fraction of one of the species is nonlinear.
The direction of the shift is dependent on the
material properties of the species 
and on the range of volume fraction,
so that the nature of the temperature shift is
more complicated than that in polymer solutions~\cite{Onuki,tao}.
Physically, this feature of polymer blends can be attributed to the
fact that both species have contributions to the elastic
free energy.

\vspace{0.5cm}
 
\begin{center}
ACKNOWLEDGMENTS
\end{center}
 
We wish to thank Professor T. Ohta for his interest and many useful
discussions.
We are grateful for the support of ONR, through grants N00014-91-J-1363
(A.C.B.) and NSF, through grant DMR-92-17935 (D.J.) 
and DMR-9709101 (A. C. B).

\end{document}